\documentclass[prl,twocolumn,showpacs,aps,superscriptaddress]{revtex4-1}
 
\usepackage{amsmath}
\usepackage{amssymb}
\usepackage{amsmath}
\usepackage{txfonts}
\usepackage{graphicx}
\usepackage{epsfig}
\usepackage[T1]{fontenc}
\usepackage[latin9]{inputenc}
\usepackage{times}
\usepackage{color} 
\usepackage{xspace}
\usepackage{amssymb,amsmath}
\usepackage{amsbsy}
\usepackage{braket}
\usepackage{tabularx}
\usepackage{mathtools}          

\DeclareMathOperator{\sech}{sech}

\usepackage{ifpdf}
\ifpdf
\usepackage{epstopdf}
\fi

\usepackage[unicode]{hyperref}
\hypersetup{
	pdftitle={spinnorBEC},
	pdfauthor={Yu},
	pdfsubject={Vortexlines},
	colorlinks=true,
	linkcolor=blue,
	citecolor=blue,
	filecolor=black,
	urlcolor=blue
}
\def\urlprefix{}
\def\url#1{}

\usepackage{bm}
\usepackage{color}

\newcommand{\be}{\begin{equation}}

\newcommand{\ee}{\end{equation}}

\newcommand{\bea}{\begin{eqnarray}}

\newcommand{\eea}{\end{eqnarray}}

\newcommand{\nn}{\nonumber }

\newcommand\smallO{
	\mathchoice
	{{\scriptstyle\mathcal{O}}}
	{{\scriptstyle\mathcal{O}}}
	{{\scriptscriptstyle\mathcal{O}}}
	{\scalebox{.7}{$\scriptscriptstyle\mathcal{O}$}}
}

\renewcommand{\Re}{\operatorname{Re}}
\renewcommand{\Im}{\operatorname{Im}}

\begin{document}


\title{Propagating Ferrodark Solitons in a Superfluid: Exact Solutions and Anomalous Dynamics} 
%

\author{Xiaoquan Yu}
\email{xqyu@gscaep.ac.cn}
\affiliation{Graduate School of  China Academy of Engineering Physics, Beijing 100193, China}
\affiliation{Department of Physics, Centre for Quantum Science, and Dodd-Walls Centre for Photonic and Quantum Technologies, University of Otago, Dunedin, New Zealand}
\author{P.~B.~Blakie}
\affiliation{Department of Physics, Centre for Quantum Science, and Dodd-Walls Centre for Photonic and Quantum Technologies, University of Otago, Dunedin, New Zealand}

\begin{abstract}
	Exact propagating topological solitons are found in the easy-plane phase of ferromagnetic spin-1 Bose-Einstein condensates, manifesting themselves as kinks in the transverse magnetization. Propagation is only  possible when the symmetry-breaking  longitudinal  magnetic field is applied. Such solitons have two types: a low energy branch with positive inertial mass and a higher branch solution with negative inertial mass. Both types become identical at the maximum speed, a new speed bound that is different from speed limits set by the elementary excitations.  The physical mass, which accounts for the number density dip, is negative for both types. In a finite one-dimensional system subject to a linear potential, the soliton undergoes oscillations  caused by transitions between the two types occurring at the maximum speed. 
	
\end{abstract}

\maketitle

\textit{Introduction---}
The inertial mass (or effective mass) of emergent quasi-particles contains rich information on the dynamics of quantum  many body systems~\cite{mahan2013many}. 
In quantum fluids the inertial mass of a topological soliton is determined by both the kinetic and interaction  energies and is a key quantity governing its dynamics. For instance, the one-dimensional (1D) motion of a dark/grey soliton in a superfluid (bosonic or fermionic)  can be described by a Newton equation with negative inertial mass~\cite{Pitaevskii2011}, leading to oscillations in a harmonic trap~\cite{Anglin2000,Pitaevskii2004,Pitaevskii2011}.
The sign of inertial mass also signals the stability of the soliton in a system of higher than one spatial dimension. Indeed, two- or three-dimensional solitons with negative inertial mass typically decay \cite{Pitaevskiisnake2008} due to the snake instability (growth of transverse deformations)~\cite{kuznetsov1988instability,Shlyapnikov1999,footnotesnake}. It is a rather general feature for solitons in quantum fluids that the soliton energy decreases with increasing velocity, giving rise to a negative inertial mass. Relevant examples are dark/grey solitons in bosonic and fermionic quantum gases~\cite{Pitaevskii2011}, phase domain walls in binary Bose-Einstein condensates (BECs) with strong coherent coupling~\cite{Gallemi2019, MDQu2017, Sophie,footnotephasedomainwall}, magnetic solitons in both binary~\cite{MDQu2016} and anti-ferromagnetic spin-1 BECs~\cite{MSexp1,MSexp2}.  A soliton with positive inertial mass should be stable in higher dimensions  and  exhibit anomalous dynamics. 

%
%

In this Letter we report on the discovery of two types of exact topological solitons that have positive and negative inertial mass, respectively,  occurring as kinks in the transverse magnetization of a ferromagnetic spin-1 BEC.  We refer to them as ferro-dark solitons (FDSs). 
In the zero velocity limit the FDSs connect to the stationary magnetic domain walls (MDWs) recently found in Ref.~\cite{MDWYuBlair}. 
The FDSs can only propagate at a finite speed in a longitudinal magnetic field which provides a necessary condition for the motion, i.e., breaking the transverse magnetization conversation. 
In addition, the FDSs exhibit a number of other novel features different from conventional solitons.  When traveling, the transverse magnetization is always zero in the core of a FDS and hence there is no magnetic current. 
The motion arises from a coupling between the magnetization and nematic degrees of freedom caused by the magnetic field.
Interestingly, the moving speed is not limited by group velocities of elementary excitations but has a new speed bound, at which the two types of solitons become identical.  We study dynamics of the soliton in a hard-wall trapped quasi-1D system with a superimposed linear potential and find transitions between the two types via internal spin currents, leading to an oscillatory motion.  While we focus on the exactly solvable case, we would like to emphasize that FDSs exist with the characteristic features revealed by the exact solutions  in the whole easy-plane phase.

\textit{Spin-1 BECs---}
The Hamiltonian density of a spin-1 condensate reads 
\bea 
\label{Hamioltonian}
{\cal H}=  \frac{\hbar^2 \left|\nabla \psi\right|^2 }{2M} +\frac{g_n}{2} |\psi^{\dag}\psi|^2+\frac{g_s}{2} |\psi^{\dag} \mathbf{S} \psi|^2 +q \psi^{\dag} S^2_z \psi,
\eea
where the three-component wavefunction $\psi=(\psi_{+1},\psi_{0},\psi_{-1})^{T}$ describes the atomic hyperfine state $\ket{F=1,m=+1,0,-1}$, $M$ is the atomic mass, $g_n>0$ is the density interaction strength, $g_s$ is the spin-dependent interaction strength,
$\mathbf{S}=(S_x,S_y,S_z)$ with $S_{j=x,y,z}$ being the spin-1 matrices~\cite{footnote-3},  and $q$ denotes the quadratic Zeeman energy. The spin-dependent interaction term allows for spin-mixing collisions between $m=0$ and $m=\pm 1$ atoms.  At the mean-field level, the  dynamics of the field $\psi$ is governed by the  Gross-Pitaevskii equations (GPEs) 
\bea
\label{spin-1GPE}
i\hbar \frac{\partial \psi_{\pm 1}}{\partial t}	&&=\left[H_0+g_s\left(n_0+n_{\pm 1}-n_{\mp 1}\right)+q \right]\psi_{\pm 1}+g_s \psi^2_0 \psi^{*}_{\mp 1} , \nn\\
i\hbar \frac{\partial \psi_0}{\partial t}	&&= \left[H_0 +g_s\left(n_{+1}+n_{-1}\right) \right]\psi_0 + 2g_s \psi^{*}_0\psi_{+1}\psi_{-1}, 
\eea 
where $H_0=-\hbar^2\nabla^2/2M +g_n n$, $n_m=|\psi_m|^2$ and $n=\sum n_m$. Spin-1 BECs support magnetic order~\cite{Ho98,OM98,Stampernatrue2006,StamperRMP,KAWAGUCHI12}, quantified by the order parameter magnetization $\mathbf{F}\equiv\psi^{\dag} \mathbf{S} \psi$. This  identifies   ferromagnetic order $|\mathbf{F}|>0$ for $g_s<0$ ($^{87}$Rb,$^7$Li) and  anti-ferromagnetic order $\mathbf{F}=0$ for $g_s>0$ ($^{23}$Na).

\textit{Quadratic Zeeman driven propagating FDSs---}
We consider a uniform ferromagnetic ($g_s<0$) spin-1 BEC with total number density $n_b$. In the presence of a uniform magnetic field  along the $z$-axis ($0<q<-2g_s n_b$)~\cite{footnoteq}, the uniform ground state with zero longitudinal magnetization ($F_z=n_{+1}^b-n_{-1}^b=0$) is transversally magnetized (easy-plane phase)~\cite{StamperRMP,KAWAGUCHI12}, characterized by the transverse magnetization  $F_{\perp}\equiv F_x+iF_y= \sqrt{8n^b_{\pm1}n^{b}_0}e^{i\tau}$, where $n^b_{\pm 1}=(1-\tilde{q})n_b/4$ and $n^b_0=n_b(1+\tilde{q})/2$ are the component densities, and $\tilde{q}\equiv-q/(2g_sn_b)$. The $\textrm{SO}(3)$ symmetry is broken by the magnetic field and the system processes the remnant $\textrm{SO}(2)$ symmetry, parameterized by the rotational angle about the $z$-axis $\tau$. 

In the following we focus on a 1D system. In the easy plane phase, exact transverse magnetic kink solutions of Eq.~\eqref{spin-1GPE} are found for a large spin-dependent interaction strength $g_s=-g_n/2$ and $0< q<-2g_sn_b$. There are two types of such traveling  kinks and the transverse magnetizations and the total number densities read 
\begin{table*}[!t]
	\fontsize{10.5}{23.2}\selectfont
	\centering 
	\begin{tabularx}{\linewidth}{p{0.78cm}|p{7.2cm}|p{7.6cm}}	
		\hline\hline 
		&type-I  & type-II  \\
		\hline
		$\psi$  & $\psi^{\rm I}_{\pm 1}(x,t)= \sqrt{n^{\pm 1}_b}\left[\alpha^{\rm I} \tanh\left(\frac{x-Vt}{\ell^{\rm I}}\right)+i\, \delta^{\rm I} \right]$ & $\psi^{\rm II}_{\pm 1}=\sqrt{n^{\pm 1}_b}\left[\beta^{\rm II}+i \kappa^{\rm II} \tanh\left(\frac{x-Vt}{\ell^{\rm II}}\right)\right]$\\ 
		& $\psi^{\rm I}_0(x,t)=\sqrt{n^{0}_b} \left[\beta ^{\rm I}+ i \, \kappa^{\rm I} \tanh\left(\frac{x-Vt}{\ell^{\rm I}}\right)\right]$ 	&  $\psi^{\rm II}_{0}=\sqrt{n^{0}_b}\left[\alpha^{\rm II} \tanh\left(\frac{x-Vt}{\ell^{\rm II}}\right) +i \, \delta^{\rm II}\right]$  \\  	
		& $\alpha^{\rm I}=-\sqrt{\frac{M V^2  (g_n n_b+q)}{q \left(q+M V^2-Q\right)}},\quad \delta^{\rm I}=\sqrt{\frac{ q-M V^2-Q}{2 q}}$ & $\alpha^{\rm II}=-\frac{\sqrt{M V^2+q-Q} \left(q-M V^2+Q\right)}{2 \sqrt{ q M V^2 \left(g_nn_b-M V^2+Q\right)}}$,\, $\delta^{\rm II}=-\sqrt{\frac{q+ M V^2-Q}{2 q}}$\\
		&$\beta^{\rm I}=\sqrt{\frac{q+M V^2+Q}{2 q}}$ , \, $\kappa^{\rm I} 
		=-\sqrt{\frac{q \left(q-Q\right)-g_n M n_b V^2}{q \left(q+M V^2-Q\right)}} $ & $\beta^{\rm II}=\sqrt{\frac{ q-M V^2+Q }{ 2 q }}$, \quad $\kappa^{\rm II}=\frac{ \left(q+M V^2-Q\right) \sqrt{q-M V^2+Q}}{2 \sqrt{q M V^2 \left(g_nn_b-M V^2+Q\right)}} $ \\ 
		\hline
		${\cal K}$ & $\frac{2 \sqrt{2} q \sqrt{n^{\pm 1}_b n^{0}_b}\,\delta^{\rm I}\, \beta^{\rm I}  }{\hbar } \sech ^2\left(\frac{x- Vt}{\ell^{\rm I}}\right)$	& $\frac{2 \sqrt{2} q \sqrt{n^{0}_bn^{\pm 1}_b}  \delta^{\rm II} \beta^{\rm II} }{\hbar }\sech ^2\left(\frac{x-Vt}{\ell^{\rm II}}\right)$  \\ 
		\hline 	
		$J^{x}_{\pm 1}$,\,$J^{x}_{0}$	& $-\frac{n^{\pm 1}_b\alpha^{\rm I} \delta^{\rm I} \hbar }{\ell^{\rm I} M}\sech^2\left(\frac{x- Vt}{\ell^{\rm I}}\right)$,\,\, $\frac{ n^{0}_b \kappa^{\rm I} \beta^{\rm I} \hbar }{\ell^{\rm I} M}\sech^2\left(\frac{x-V t}{\ell^{\rm I}}\right) $
		& $\frac{ n^{\pm 1}_b \kappa^{\rm II} \beta^{\rm II} \hbar  }{\ell^{\rm II} M} \sech^2\left(\frac{x-V t}{\ell^{\rm II}}\right)$, \, \, $-\frac{n^{0}_b \alpha^{\rm II}  \delta^{\rm II} \hbar  }{\ell^{\rm II} M} \sech^2\left(\frac{x- Vt}{\ell^{\rm II}}\right)$\\
		\hline
		$J_{\pm 1 \rightarrow 0}$	& $-\frac{4  g_s \left(\delta^{\rm I}  \kappa^{\rm I} +\beta^{\rm I} \alpha^{\rm I}\right)  \, \delta^{\rm I} \, \beta^{\rm I}  n^{\pm 1}_b  n^{0}_b }{\hbar } \tanh\left(\frac{x-V t}{\ell^{\rm I}}\right)\sech ^2\left(\frac{x-V t}{\ell^{\rm I}}\right) $ & $\frac{4 g_s \left(\kappa^{\rm II}  \delta^{\rm II} + \beta^{\rm II}   \alpha^{\rm II}\right) \delta^{\rm II}  \beta^{\rm II}n^{\pm 1}_b  n^{0}_b }{\hbar } \tanh \left(\frac{x-V t}{\ell^{\rm II}}\right) \sech^2 \left(\frac{x-Vt}{\ell^{\rm II}}\right)$ \\ 
		\hline\hline
	\end{tabularx} 
	\caption{Wavefunctions and currents of propagating FDSs in the exactly solvable regime ($g_s=-g_n/2$, $0 < q<-2g_s n_b$). The coefficients satisfy the following relations: $\kappa^{\rm I,II} \alpha^{\rm I,II}=\delta^{\rm I,II} \beta^{\rm I,II}$ and \, $ (\alpha^{\rm I,II})^2+(\delta^{\rm I,II})^2=(\beta^{\rm I,II})^2+(\kappa^{\rm I,II})^2=1$.  It is straightforward to check that stationary solutions are obtained when $V\rightarrow 0$~\cite{footnotesmallvelocity,MDWYuBlair}. Here ${\cal{K}}^2 \equiv  { \sum_{i} K_{iz} ^2}$ is $\textrm{SO(2)}$ rotationally invariant~\cite{footnotesourcecomponents}. The counter-propagating solution is $\psi^{*}(x,-t)$.   }
	\label{tableI}
\end{table*}
\bea
F^{\rm I, II}_{\perp}(x,t)
&=& -e^{i \tau } \sqrt{ n_b^2-\frac{ q^2}{g_n^2}} \tanh \left(\frac{x-Vt}{\ell^{\rm I, II}}\right),\\
n^{\rm I,II}(x,t)
&=&n_b -\frac{g_n n_b-M V^2\mp Q}{2 g_n}\sech ^2\left(\frac{x-V t}{\ell^{\rm I,II}}\right),
\eea
where $V$ is the moving velocity,
\bea
\ell^{\rm I,II}= \sqrt{\frac{2\hbar ^2}{M \left(g_n n_b-M V^2\mp Q\right)}}, 
\label{length}
\eea
and
\bea
Q=\sqrt{M^2 V^4+q^2-2 g_n M n_b V^2}.
\eea
The above kink solutions are of Ising-type and connect regions transversely magnetized in opposite directions~\cite{footenotexcitation}.
Hereafter we refer to them as ferro-dark solitons (FDSs) and the minus (plus) sign in front of $Q$ specifies type-I (II) FDS.
Unless specified, we choose $\tau=0$  for convenience. 
At the core, the transverse magnetization $F_{\perp}$ is zero while the component densities $n_{0,\pm1}$ do not vanish for finite velocity $V$.
The corresponding wavefunctions at the exactly solvable region are shown in Table~\ref{tableI}.  Recently a $^7$Li spin-1 BEC has been prepared in the strong spin interacting regime close to the exactly solvable point~\cite{Huh2020a}.

\begin{figure}[htp] 
	\centering
	\includegraphics[width=0.46\textwidth,,height=180pt]{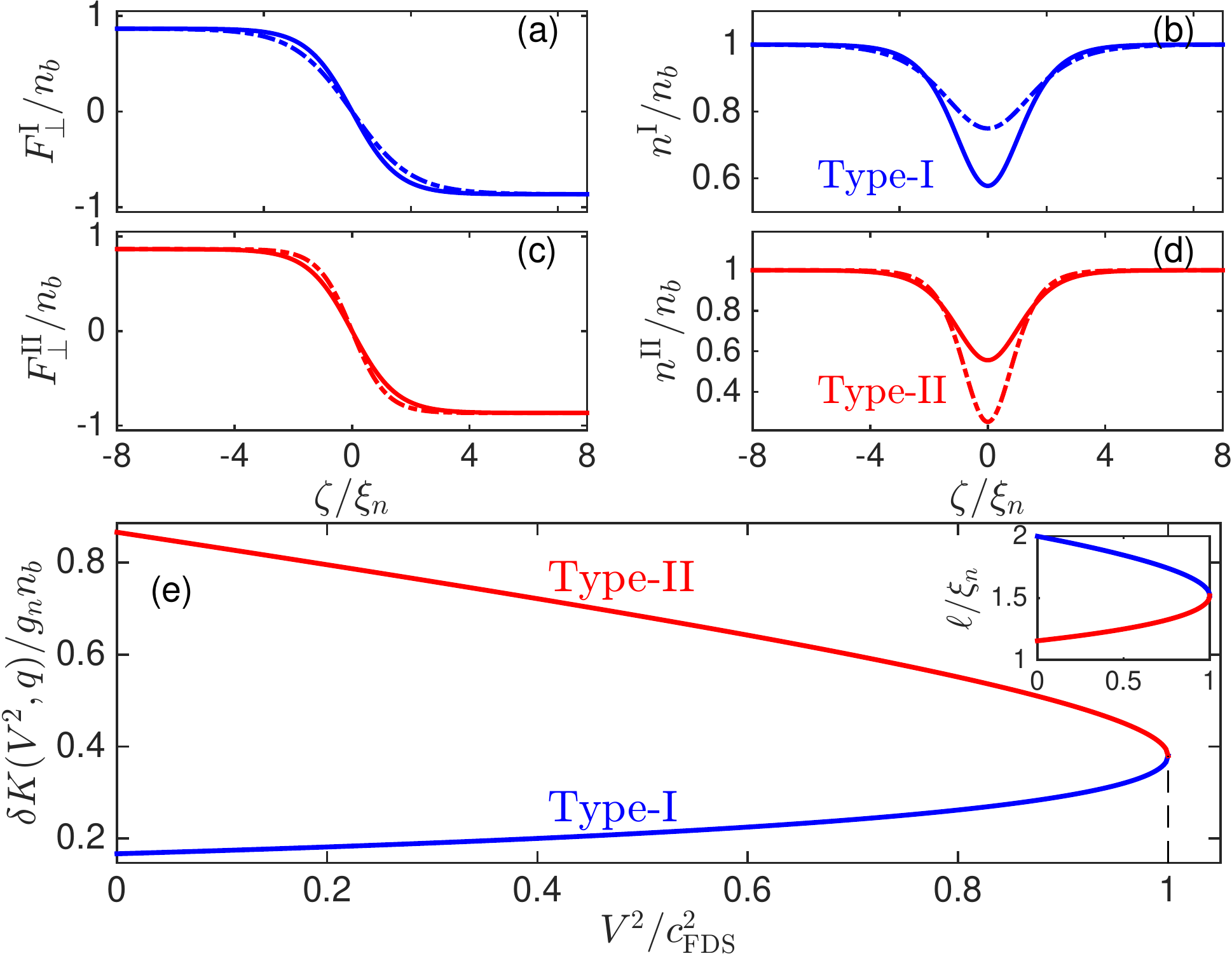}
	\caption{(a)-(d) Transverse magnetizations and densities of FDSs at $g_s=-g_n/2$ and $\tilde{q}=0.5$ for different velocities: $V/c_{\rm FDS}=1$ (solid line:);  $V/c_{\rm FDS}=0.1$ (dashed line). (e) Excitation energies of FDSs as functions of $V^2$ evaluated from Eq.~\eqref{energyI} and Eq.~\eqref{energyII} at $\tilde{q}=0.5$.  Here $\zeta=x-Vt$.  The inset shows  widths of FDSs, where the $x$-axis is the same as in (e). } 
	\label{f:movingMDW}
\end{figure}

The inequality $Q^2\geq 0$ gives rise to the upper bound of the traveling speed~\cite{footnote1}
\bea
V\leq\sqrt{\frac{g_n n_b}{M}}\sqrt{1-\sqrt{1-\left(\frac{q}{g_n n_b}\right)^2}}\equiv  c_{\rm{FDS}}.
\label{vbound}
\eea 
The speed bound Eq.~\eqref{vbound} is markedly different from the group velocities of low-lying elementary excitations which normally set the speed limits~\cite{pitaevskii2016bose}. In the easy-plane phase, the gap-less branches of the elementary excitations involve  spin waves of magnetization $\mathbf{F}$ (dominantly) and mixed waves of $F_{\perp}$ and $n$, with group velocities at long wavelengths $c_{\rm m}=\sqrt{q/(2 M)}$ and $c_{\rm mp}=\sqrt{ n_b (g_n+g_s)/M}$, respectively~\cite{SM}.  Strikingly, for $1>q/g_n n_b> \sqrt{3}/2$, $c_{\rm{FDS}}>c_{\rm mp}>c_{\rm m}$,  implying that the FDSs can travel with speed greater than $c_{\rm m}$ and $c_{\rm mp}$. This can happen because a propagating FDS does not involve magnetic currents (see below).
Another conspicuous feature is that the soliton profile does not vanish at $V=c_{\rm FDS}$ (see Fig.~\ref{f:movingMDW}). 
The velocity of grey solitons in scalar BECs is bounded by the speed of sound, and at this velocity the soliton disappears~\cite{pitaevskii2016bose}.
At the transition point $q=g_n n_b$, the easy-plane phase becomes unstable, signalled by the divergence of $\ell^{\rm I}$.

Similar to scalar gray solitons, the density dip of the type-II FDS becomes shallower for greater velocities [Fig.~\ref{f:movingMDW}(d)]. 
In contrast, for the type-I FDS the density dip behaves anomalously and deepens with increasing velocity [Fig.~\ref{f:movingMDW}(b)]. 
Crucially, at the maximum velocity $V=c_{\rm FDS}$, $Q=0$ and the two types of FDSs become identical upon a $\textrm{U}(1)$ gauge transformation,
namely $\psi^{\rm I}(x,t)=i\psi^{\rm II}(x,t)$ (see Table~\ref{tableI}).

When $q\rightarrow 0$, $c_{\rm FDS} \rightarrow 0$, implying that the propagation is prohibited in the absence of a magnetic field, where the conservation law of magnetization is restored. In this limit, the two types become degenerate and are related via a $\textrm{SO}(3)$ spin-rotation~\cite{MDWYuBlair}. 
Clearly, a magnetic field does not automatically induce motion. At $V=0$ the FDSs recover stationary MDWs at finite $q$~\cite{MDWYuBlair,footnoteboost}.

\textit{Currents---}
Moving FDSs  involve nematic degrees of freedom and internal spin currents.  Since the magnetization is zero at the core of a moving FDS, there is no magnetic current, i.e., $\mathbf{J}^{F}_i\equiv\hbar/(2M i) (\psi^{\dag} S_i \nabla \psi-\rm{H.c.})=0$. According to the continuity equation 
\bea
\frac{\partial F_i}{\partial t}+\nabla \cdot \mathbf{J}^{F}_i=K_{iz},
\eea 
the time  evolution of magnetic domains  enclosed by the MDWs is governed by the source term
$K_{iz}=(2 q/\hbar)\hat{K}_{iz}$~\cite{currents,footnotesource},
where $\hat{K}_{iz}=\sum_{k}\epsilon_{izk}N_{zk}$, $N_{ij}=\psi^{\dag}\hat{N}_{ij}\psi$ is the nematic tensor, $\hat{N}_{ij}=(S_iS_j+S_jS_i)/2$
and $i,j\in\{x,y,z\}$.
For propagating FDSs $\hat{K}_{iz}\neq 0$ and $\hat{K}_{iz} \rightarrow 0$ as $V\rightarrow 0$. At $q=0$, $K_{iz}=0$, and FDSs must stay still. 

The continuity equations for particle number in each spin state read
$\partial n_{\pm 1}/\partial t+\nabla \cdot \mathbf{J}_{\pm 1}+J_{\pm 1\rightarrow 0}=0$, and 
	$\partial n_0/\partial t+\nabla \cdot \mathbf{J}_0+\sum_{m=-1,+1}J_{0\rightarrow m}=0$,
where $\mathbf{J}_{\pm 1,0}=\hbar/(2M i)(\psi^{*}_{\pm 1,0}\nabla \psi_{\pm 1,0}-\rm{H.c.})$
are the component number current densities~\cite{footnotenumbercurrents}, and 
\bea
J_{\pm 1\rightarrow 0}=-J_{0 \rightarrow \pm 1}=\frac{g_s}{\hbar i}\left[(\psi^*_0)^2\psi_{-1} \psi_{+1}-\rm{H.c.}\right] 
\label{internalcurrent}
\eea 
are the internal spin currents, reflecting the internal coherent spin exchange dynamics: $\ket{00} \leftrightarrow\ket{+1}\ket{-1}$~\cite{Ho98,OM98,Stampernatrue2006}.  Rewriting Eq.~\eqref{internalcurrent} in terms of wavefunction phases ($\theta_{\pm1,0}$) and densities, we obtain  $J_{\pm 1\rightarrow 0}=(2n_0n_{\pm1}g_s/\hbar)\sin[2(\theta_{\pm 1}-\theta_0)]$ which suggests an analogy to Josephson currents~\cite{barone1982physics,footnotecurrents}. It is important to note that these built-in currents are invariant under $\textrm{SO}(2)$ rotations ($ e^{-i \tau S_z}$). 
Table~\ref{tableI} shows the expressions of currents at the exactly solvable  point.  Interestingly,  $J^{x}_{\pm 1}$ and $J^{x}_{0}$  have opposite signs and $\int dx \,J_{\pm1 \rightarrow 0}=0$, forming a Josephson vortex-like structure near the core of a FDS. 

\textit{Excitation energy and inertial mass---}
The excitation energy of  FDSs can be obtained  by evaluating the difference of grand cannonical energies 
$\delta K=K_{\rm FDS}-K_g$, where $K_{\rm FDS}=\int dx \, ({\mathcal H}[\psi]- \mu n )$, $K_g=\int dx \, ({\mathcal H}[\psi_g]- \mu n_b )$, $\psi_{g}$ is the ground state wavefunction and $\mu=(g_n+g_s)n_b+q/2$ is the chemical potential.
For type-I FDSs,  we obtain
\bea
\delta K^{\rm I}(q,V^2)= \frac{\sqrt{2} \hbar  \left(g_n n_b-M V^2-Q\right)^{3/2}}{3 g_n \sqrt{M}} 
\label{energyI}.
\eea
Expanding Eq.~\eqref{energyI} around $V=0$, we have 
$\delta K^{\rm I}(q,V^2)= \delta K^{\rm I} (q,0)+ M^{\rm I}_{\rm in} V^2/2+ \smallO (V^2)$
where $\delta K^{\rm I}(q,0)=\sqrt{2} \hbar (g_n n_b-q)^{3/2}/(3 g_n \sqrt{M})$
and the inertial mass  is
\bea
\hspace{-8.6pt}M^{\rm I}_{{\rm in}}&\equiv&2\frac{\partial \delta K^{\rm I}}{\partial V^2}\biggr|_{V=0}=  \frac{\sqrt{2M} \hbar  (g_n n_b-q)^{3/2}}{ g_n q}>0.
\eea
As $q\rightarrow 0$, $M^{\rm I}_{\rm in}\rightarrow +\infty$ and the FDS becomes infinitely heavy, consistent with the absence of propagation at zero magnetic field due to the conservation of magnetization~\cite{MDWYuBlair}.   In contrast to the normal behavior of grey solitons, the excitation energy ($\delta K^{\rm I}$) of the type-I FDS increases monotonically with increasing $V^2$ [Fig.~\ref{f:movingMDW}(e)],
in accordance with the anomalous behavior of the density [Fig.~\ref{f:movingMDW}(b)].  It is worth noting that here every component density has a dip (see Table~\ref{tableI}) and the inertial mass of type-I FDSs being positive is a highly non-trivial nonlinear effect. Following conventional arguments \cite{Pitaevskiisnake2008} the positive inertial mass  explains the stability of MDWs against transverse snake perturbations in 2D~\cite{MDWYuBlair}.

The physical mass is defined as $M_{\rm phy}\equiv M\delta N$, where $\delta N=\int dx \, [n(x)-n_b]$. 
For type-I FDSs,  we obtain $M^{\rm I}_{\rm phy}= -2 \hbar^2 /(g_n  \ell^{\rm I}) <0$. In the presence of an external potential $U$, a soliton with negative physical mass experiences an effective force from the surrounding liquid pointing in the opposite direction to  $-\nabla U$ (similar to buoyant force) . For a scalar grey soliton the inertial and the physical masses are both negative and it exhibits normal particle-like behavior, e.g., oscillations in a harmonic potential~\cite{Pitaevskii2004,Pitaevskii2011}. 
Whereas a type-I FDS in a harmonic potential would be expelled, i.e., moves away from the potential minimum.

The excitation energy of the type-II FDS is  
\bea
\delta K^{\rm II}(q,V^2)= \frac{\sqrt{2} \hbar  \left(g_n n_b-M V^2+Q\right)^{3/2}}{3 g_n \sqrt{M}}
\label{energyII}
\eea
with $\partial \delta K^{\rm II} / \partial V^2 <0$ [Fig.~\ref{f:movingMDW}(e)]. 
Expansion of Eq.~\eqref{energyII} leads to
$\delta K^{\rm II}(q,V^2)= \delta K^{\rm II}(q,0)+ M^{\rm II}_{\rm in} V^2/2+ \smallO (V^2)$,
where $\delta K^{\rm II} (q,0)=\sqrt{2} \hbar  (g_n n_b+q)^{3/2}/(3 g_n \sqrt{M})$
and the inertial mass 
\bea
M^{\rm II}_{\rm in} \equiv 2\frac{\partial \delta K^{\rm II}}{\partial V^2}\biggr|_{V=0}=-\frac{\sqrt{2M} \hbar  (g_n n_b+q)^{3/2}}{g_n q}<0.
\eea
Consistently,  $M^{\rm II}_{\rm in}\rightarrow -M^{\rm I}_{\rm in}\rightarrow -\infty$ as $q\rightarrow 0$.  
The physical mass  $M^{\rm II}_{\rm phy}= -2\hbar^2/(g_n \ell^{\rm II})<0$.  Thus, the inertial and physical mass of the type-II FDS is similar to those of ordinary grey/dark solitons. Excitation energies of type-I and type-II FDSs coincide  smoothly  at the maximum speed [Fig.~\ref{f:movingMDW}(e)], making transitions between the two types of FDSs possible under certain circumstances.

\begin{figure}[htp] 
	\centering
	\includegraphics[width=0.47\textwidth]{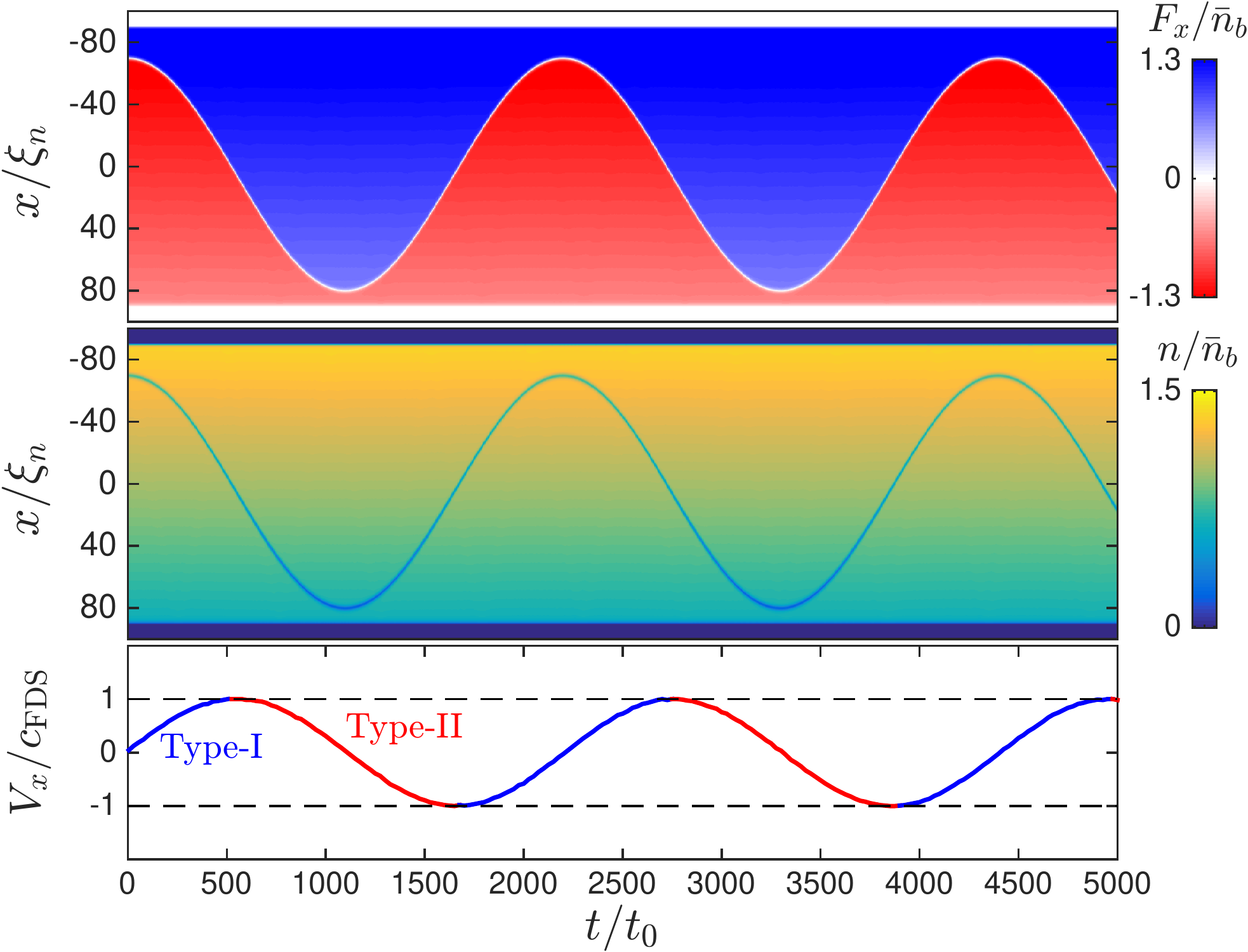}
	\caption{Oscillations of a FDS in a hard-wall trapped spin-1 BEC with a superimposed linear potential~\cite{footnotelinearpotential}. The system size is  $200\xi_n$,  $g_s/g_n=-1/2$ and $\tilde{q}=q/(-2g_s \bar n_b)=0.3$. Here ${\bar n_b}$ is the average density, $t_0=\hbar/g_n {\bar n_b}$ and $\xi_n=\hbar/\sqrt{M g_n \bar{n}_b}$ is the density healing length. Upper and middle panels show spin and density dynamics of a  FDS, respectively. The transverse magnetization is always zero at the core (see also Fig.~S3~\cite{SM}) and the topological characteristic, i.e., the sign change of $F_{x}$ is kept.  Bottom panel shows the  velocity of the FDS as a function of time, obtained by taking the derivative of its position with respect to time. The slope refers to the acceleration of the FDS and indicates the sign of the inertial mass (positive: blue; negative: red). The transition between type-I and type-II FDSs occurs when the slope changes sign at the maximum speed. Here $c_{\rm FDS}$ is the local speed limit for the (background) density at the position where $dV_x/dt$ changes sign.  } 
	\label{f:oscillation}
\end{figure}

\textit{Oscillations between type-I and type-II FDSs ---}
As discussed earlier the FDS does not vanish as $V\to c_\text{FDS}$, so a natural question is what will happen if it is further accelerated? 
Let us consider a hard-wall trapped quasi-1D spin-1 BEC subjected to a linear potential whose gradient is along the positive $x$-axis.  
A  $V=0$ type-I FDS is initially placed near the left end of the system, and the later dynamic shows, surprisingly,  a periodic motion.  
The FDS accelerates until it reaches the maximum speed (the local value of $c_{\rm FDS}$~\cite{footnotelocalspeedlimit}) at which point it smoothly transforms into a type-II FDS. 
Due to the sign change of the inertial mass (or more generally $\partial \delta K^{\rm I} / \partial V^2 >0\rightarrow  \partial \delta K^{\rm II} / \partial V^2<0$), it starts to accelerate in the opposite direction.
After reaching the turning point,  the FDS starts to move to the left.  It converts back to the type-I FDS and experiences positive acceleration again when gaining the maximum speed. Later it returns to the initial configuration.
Note that during the motion there is no sign change of the physical mass.  Numerical simulations show that this process continues without decay (see Fig.~\ref{f:oscillation} and a movie~\cite{SM})

\begin{figure}[htp] 
	\centering
	\includegraphics[width=0.475\textwidth]{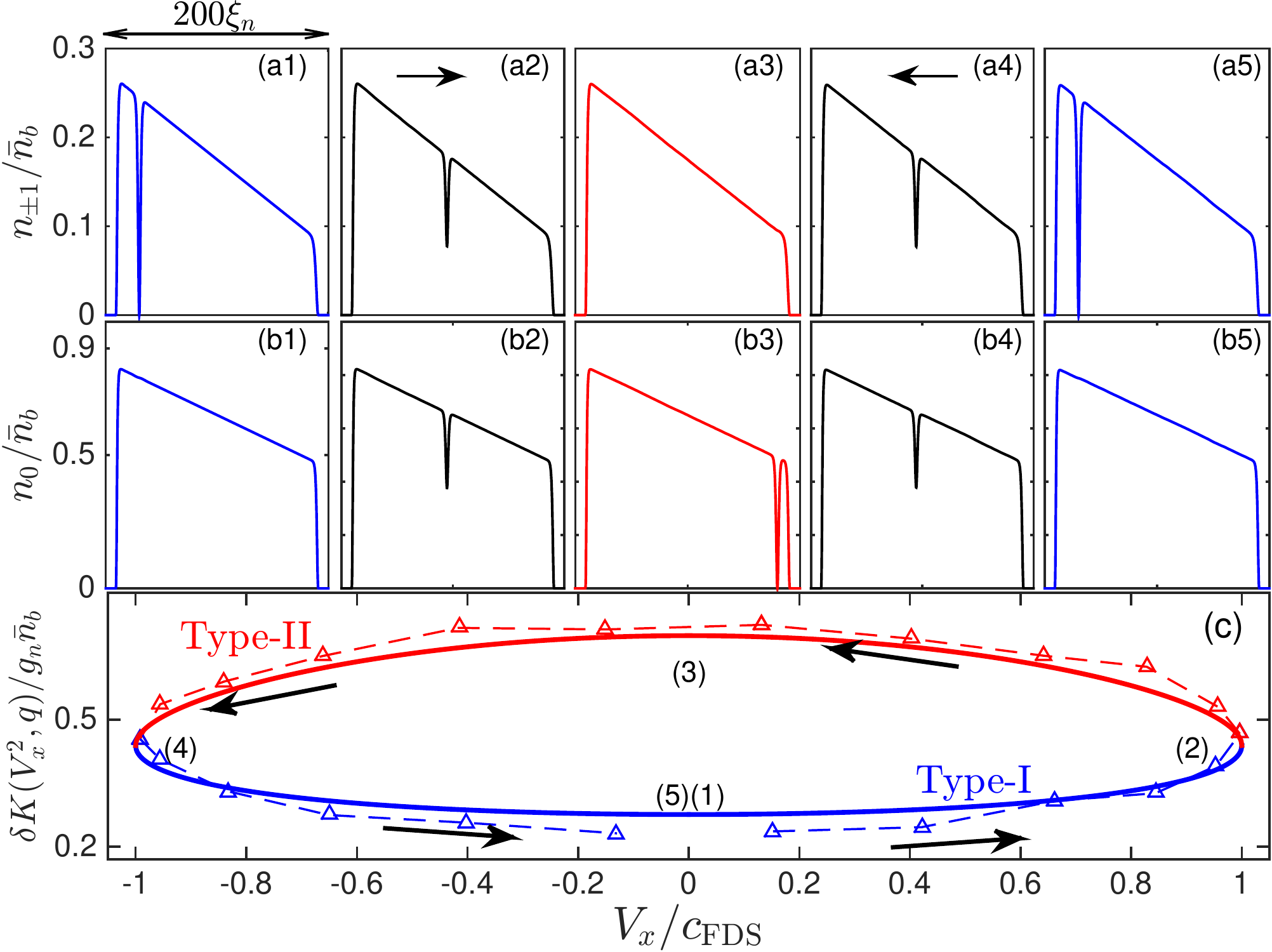}
	\caption{Internal oscillations between  $m=\pm1$ and $m=0$ spin states and the excitation energy for one complete cycle of the motion described in Fig.~\ref{f:oscillation}. The black arrows specify the evolution direction. (a1)-(b5) show component densities of the initial state (type-I FDS with zero velocity) [blue], at the maximum velocity [black], at the turning point (type-II FDS with zero velocity)[red], at the negative maximum speed [black], and of the final state (returning the initial state) [blue], respectively. (c) shows analytical predictions (solid lines) for $n_b=\bar{n}_b$ vs. numerical results (markers) for the mapped uniform system with the same density (see main text). Number labels indicate the stages corresponding to those showing in the upper panels. Note that the total energy~\cite{footnotetotalenergy} is conserved.
	} 
	\label{f:internaloscillationandenergy}
\end{figure}

During the motion the total number density profile of the soliton has only minor changes with respect to the local background density (see Fig.~\ref{f:oscillation} and Fig.~S3~\cite{SM}). However internal oscillations (driven by the gradient of the external potential) between $m=\pm 1$ and $m=0$ spin states near the core take place though the internal currents $J_{\pm 1\leftrightarrow 0}$ (Fig.~\ref{f:internaloscillationandenergy} and Fig.~S4~\cite{SM}), inducing transitions between type-I and type-II FDSs. 
Accounting for the varying density and the potential energy,  we map the FDS energy $\delta K$ extracted from the simulation to its corresponding values for a uniform system~\cite{SM}, and find that it oscillates between lower branch (type-I) and higher branch (type-II) (Fig.~\ref{f:internaloscillationandenergy}(c)), as predicted.  Here we adopt linear potentials to give a transparent illustration of the FDS dynamics. Transition between the two types FDSs can take place in other situations when the maximum speed is reached.

It should be emphasized that away from the exact solvable parameter region ($g_s=-g_n/2$, $0<q<-2g_s n_b$) the characteristic features of the oscillating dynamics hold in general (Fig.~S5~\cite{SM}).
Such an oscillation is a nonlinear phenomenon and is a result of a combination of internal spin currents induced by spin-dependent interactions, the external potential and two types of solitons being smoothly connected at the maximum speed. It occurs in a system without built-in periodicity and is distinct from the celebrated phenomenon of Bloch oscillations where the key ingredient is the presence of a band structure.

\textit{Conclusion---}  We discover a propagating magnetic kink corresponding to a topological soliton with negative physical mass and positive inertial mass in the easy-plane phase of a ferromagnetic spin-1 BEC. It can convert to its higher energy counterpart with negative physical and inertial mass at a novel maximum speed that can be greater than the  group velocities of elementary excitations which normally set the speed limits. 	The transition between the two types induces oscillations in a linear potential~\cite{footnoteoscillations}. 
Our findings open up a possibility to explore novel domain wall/soliton dynamics and could be highly relevant to out-of-equilibrium quench dynamics in 1D ferromagnetic superfluids~\cite{Prufer2018a,Gasenzer2019}.
Advances in engineering optical potential~\cite{Dalibard2015,Gauthier16,Semeghini2018} and nondestructive spin-sensitive imaging methods~\cite{Higbie2005,Semeghini2018,Kunkel2019} open the possibility of experimental investigations of the ferrodark soliton dynamics.

\textit{Acknowledgment---}
We thank M. Antonio, L. Qiao, D. Baillie and Y. Yang for useful discussions. We particularly thank J. N. BiGuo for pointing out that the energy expression of type-I FDSs can be simplified to the current form. X.Y. acknowledges  support from NSAF (No. U1930403) and NSFC (No. 12175215). P.B.B acknowledges support from the Marsden Fund of the Royal Society of New Zealand.


\begin{thebibliography}{51}
\bibitem [{foo({\natexlab{h}})}]{footnote1}%
	\BibitemOpen
	\href@noop {} {\emph {\bibinfo {title} {\rm {The other branch is
					$y\geq\sqrt{1+\sqrt{1-(q/g_n n_b)^2}}$, where $y=V/\sqrt{g_n n_b/M}$. However
					the condition $1-y-\sqrt{y^2 +(q/g_n n_b)^2-2 y}>0$ cannot be satisfied if
					$y>1$. Hence this branch is not a solution. Other constraints $q-Q-M V^2>0$,
					$q+M V^2-Q>0$, and $q \left(Q-q\right)+g_n M n_b V^2<0$ can be reduced to
					$q-Q-M V^2>0$ which is always satisfied as long as $q/g_n n_b<1$ and $M
					V^2/g_n n_b<q/g_n n_b$. Since $\sqrt{1-\sqrt{1-(q/g_n n_b)^2}}<q/g_n n_b$,
					these constraints are automatically satisfied.}}}}\BibitemShut {Stop}%
\end{thebibliography}

%

\pagebreak
\widetext
\begin{center}
	\textbf{\large Supplemental Material for `` Propagating ferrodark solitons in a superfluid: Exact solutions and Anomalous dynamics ''}
\end{center}

\setcounter{equation}{0}
\setcounter{figure}{0}
\setcounter{table}{0}
\setcounter{page}{1}
\makeatletter
\renewcommand{\theequation}{S\arabic{equation}}
\renewcommand{\thefigure}{S\arabic{figure}}
\renewcommand{\bibnumfmt}[1]{[S#1]}
\renewcommand{\citenumfont}[1]{S#1}

\section{Elementary excitations in the easy-plane phase}
Let us denote $\psi_g$ as the ground state wavefunction in the easy-plane phase ($0<q<-2g_sn_b$). Substituting the perturbed  wavefunction $\psi=\psi_g+\delta \psi$ with $\delta \psi=\epsilon[u(x)e^{-i\omega t}+v^{*}(x)e^{i \omega^{*} t}]$ into 1D Gross-Pitaevskii equations (Eq.~(2) in the main text) and keeping the leading order terms, we obtain the bosonic Bogoliubov-de Gennes (BdG) equations 
\begin{figure}[htp] 
	\centering
	\includegraphics[width=0.37\textwidth]{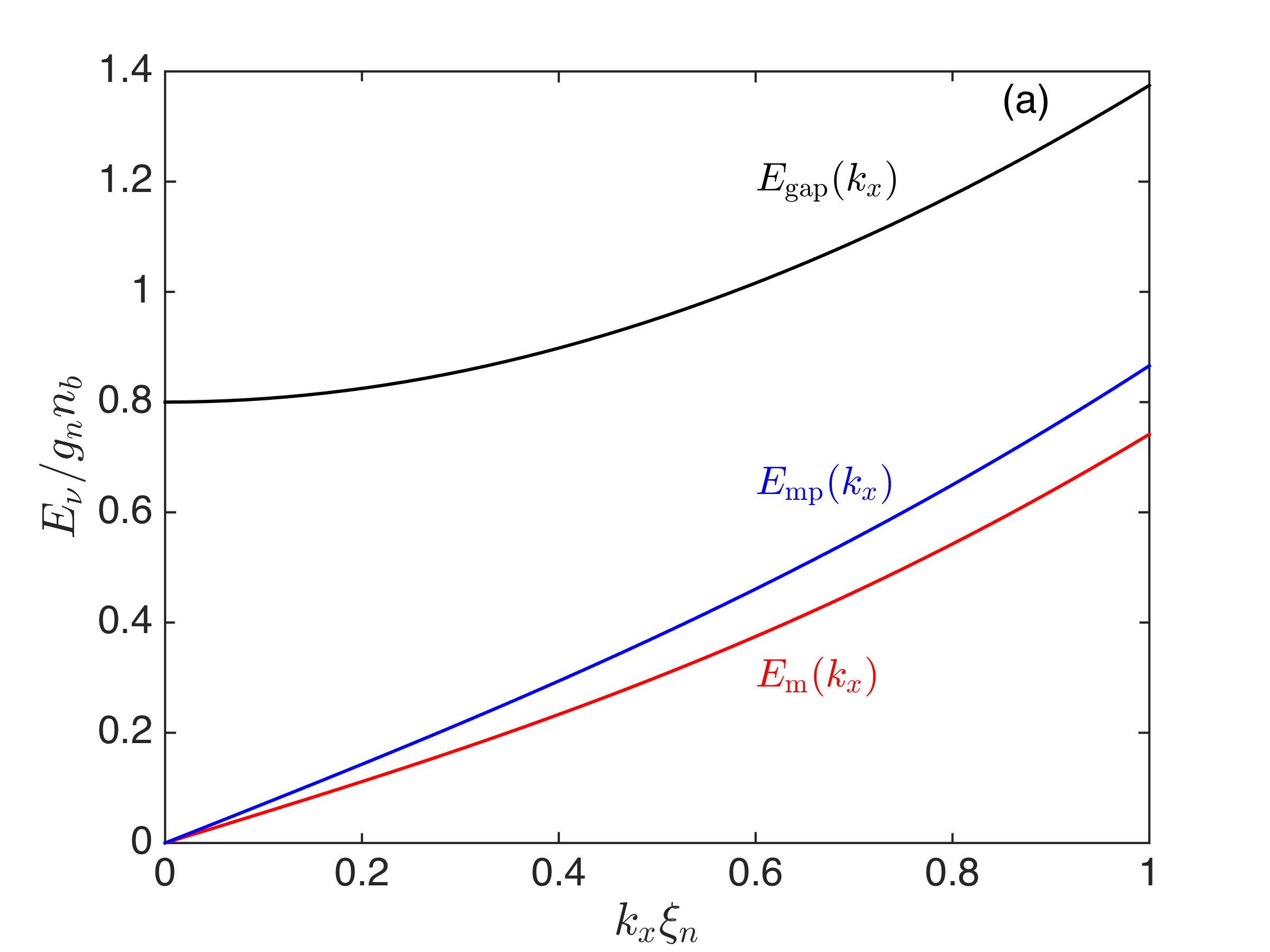}
	\includegraphics[width=0.37\textwidth]{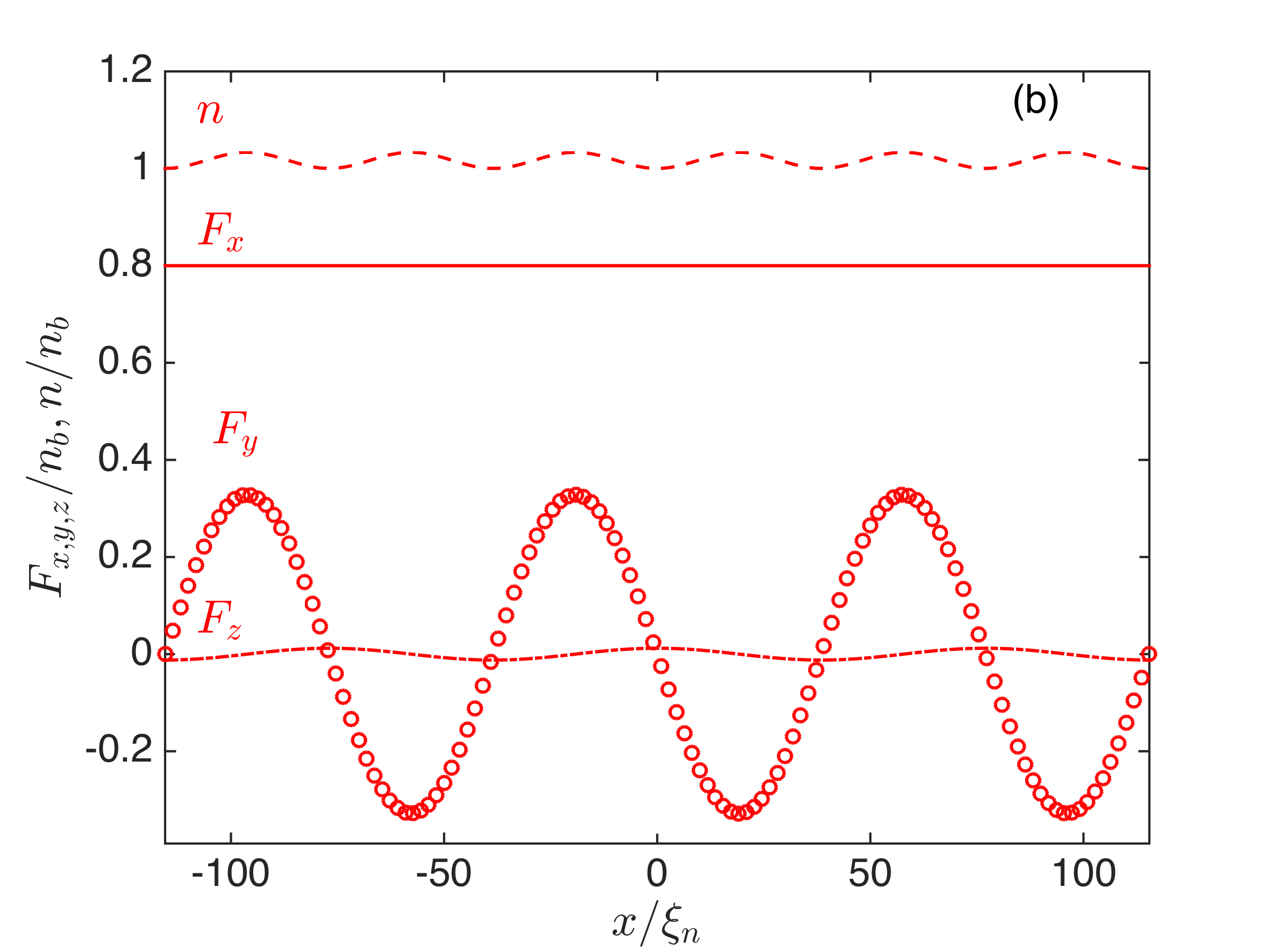}
	\includegraphics[width=0.37\textwidth]{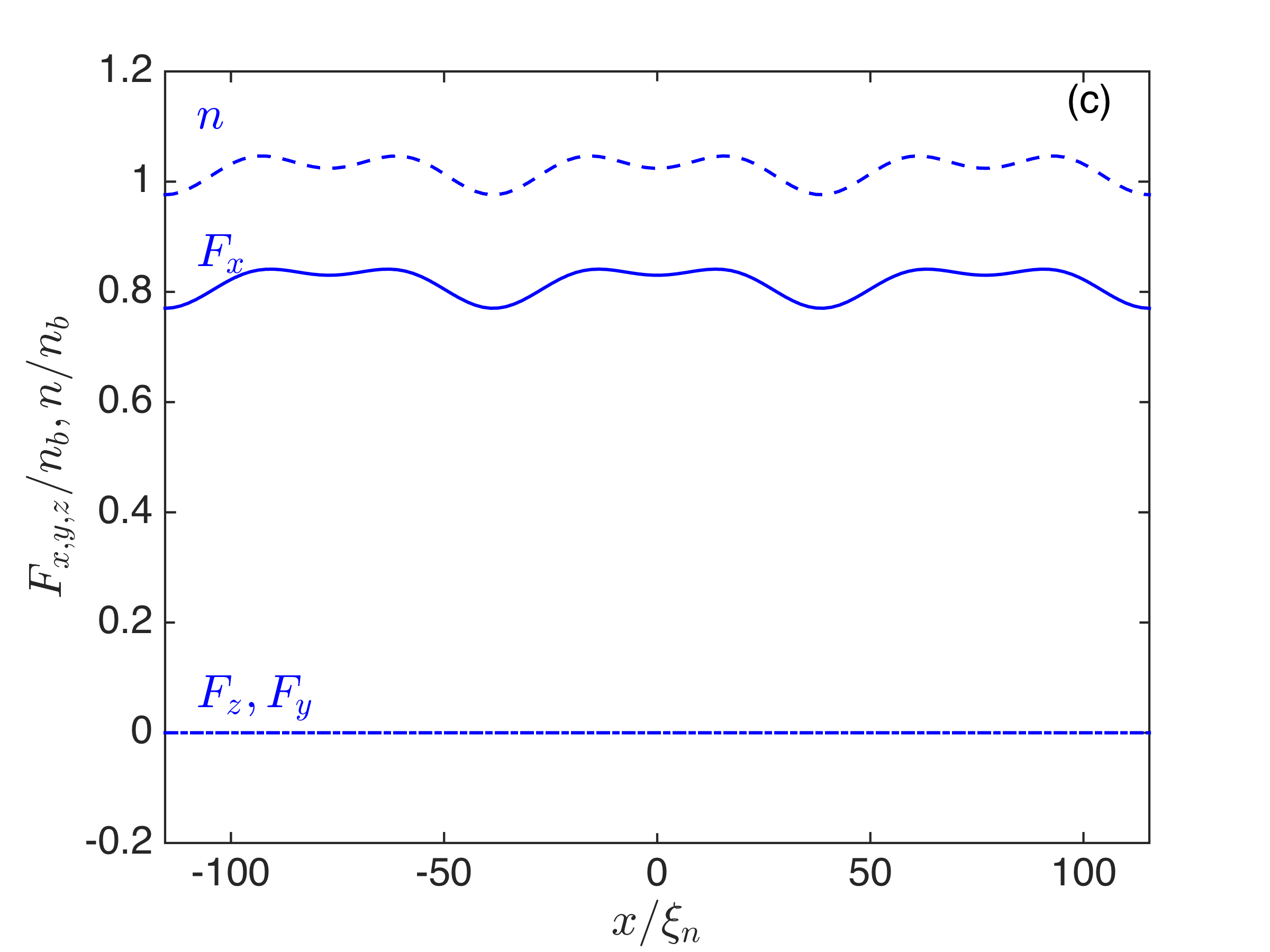}
	\includegraphics[width=0.37\textwidth]{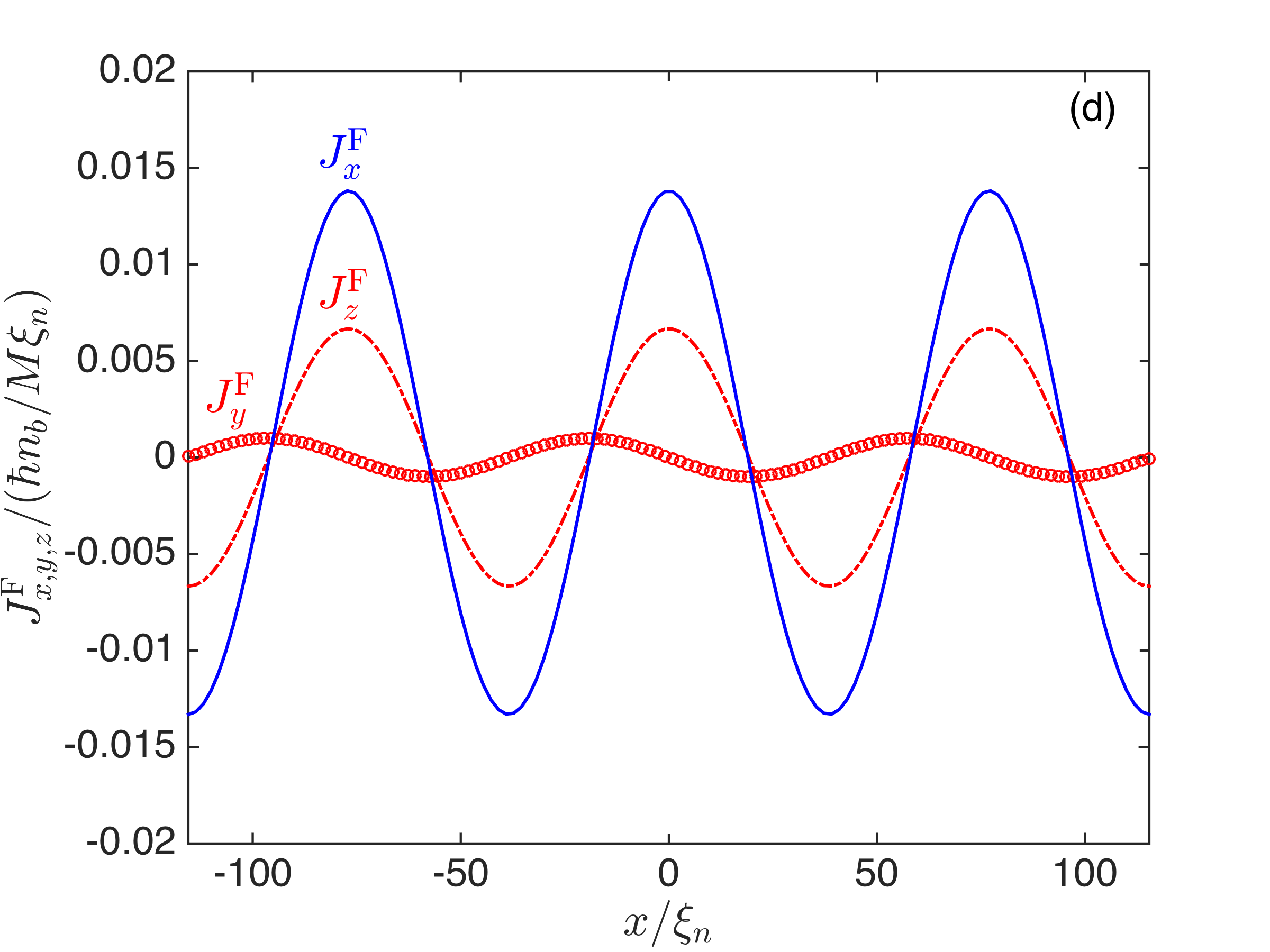}
	\caption{Elementary excitations in the easy-plane phase for  $g_s=-g_n/2$ and $\tilde{q}=q/(-2g_s n_b)=0.6$.  Here $\xi_n=\hbar/\sqrt{M g_n n_b}$ is the density healing length. We choose $\tau=0$ and hence the ground state is $\psi_{g}=(\sqrt{(1-\tilde{q})n_b/4},\sqrt{n_b(1+\tilde{q})/2},\sqrt{(1-\tilde{q})n_b/4})^{\rm T}$, where $\tilde{q}=-q/(2g_sn_b)$. The corresponding magnetization reads $F_x[\psi_g]= n_b\sqrt{1-\tilde{q}^2}$, $F_y[\psi_g]=0$ and $F_z[\psi_g]=0$. (a) shows the spectrum of the excitations.  (b), (c) show perturbed observables $F_x[\psi]$, $F_y[\psi]$, $F_z[\psi]$, and $n[\psi]$ for magnon [dominantly] (red) and magnon-phonon (blue) 	branch excitations, respectively, 
		with $k_x=0.0816/ \xi_n$.  In (c) the second order contribution is considerable, leading to waves of mixed wave lengths ($2\pi/k_x$ and $\pi/k_x$). (d) shows the non-zero magnetic currents $J^{F}_i\equiv\hbar/(2M i) (\psi^{\dag} S_i \nabla \psi-\rm{h.c.})$ for these excitations.  Here $t=0$ and $\epsilon=0.05$. }
	\label{f:BdGspectrum}
\end{figure}
\bea
E \left( {\begin{array}{cc}
		u\\
		v\\
\end{array} } \right)=
\left( {\begin{array}{cc}
		{\cal L}_{\rm GP}+X-\mu & \Delta \\
		-\Delta^{*} & -({\cal L}_{\rm GP}+X-\mu)^{*} \\
\end{array} } \right)
\left( {\begin{array}{cc}
		u\\
		v\\
\end{array} } \right),
\label{BdG}
\eea
where $\epsilon$ is a dimensionless small number, $E=\hbar \omega$, ${\cal L}_{\rm GP}\equiv-\hbar^2\nabla^2/2M +g_n \psi^{\dagger}_g\psi_g+g_s\sum^{3}_{i=1}\psi^{\dagger}_gS_i\psi_gS_i+q m^2$, $X\equiv g_s\sum^{3}_{i=1} S_i\psi_g\psi_g^{\dagger}S_i+ g_n \psi_{g}\psi^{\dagger}_g $, $\Delta\equiv g_n \psi_g \psi^{T}_g+g_s \sum^3_{i=1} (S_i \psi_g)(S_i\psi_g)^{T}$ and $\mu=(g_n+g_s)n_b+q/2$ with $n_b=|\psi_g|^2$. Note that ${\cal L}_{\rm GP} \psi_g=\mu \psi_g$. Since the system has  translational symmetry, it is natural to parameterize the perturbations according to the wave-vector $k_x$:  $u(x)=u e^{ik_x x}$ and $v(x)=v e^{ik_x x}$. Solving Eq.~\eqref{BdG}, we obtain  
\bea
E_{\rm m}(k)&=&\pm \frac{\hbar  \sqrt{(k_x^2 \hbar ^2+2 M q) k^2_x}}{2 M},\\
E_{\rm mp}(k)&=&\pm \frac{\sqrt{g_s \left(g_s \left(2 g_n k_x^2 M n_b \hbar ^2+k_x^4 \hbar ^4-2 M^2 q^2\right)+8 g_s^3 M^2 n_b^2-2 g_s^2 k_x^2 M n_b \hbar ^2+2 M \Gamma_k\right)}}{-2 g_s M}, \\
E_{\rm gap}(k)&=&\pm \frac{\sqrt{g_s \left(g_s \left(2 g_n k_x^2 M n_b \hbar ^2+k_x^4 \hbar ^4-2 M^2 q^2\right)+8 g_s^3 M^2 n_b^2-2 g_s^2 k_x^2 M n_b \hbar ^2-2 M \Gamma_k\right)}}{-2 g_s M},
\eea
where $\Gamma_k=\sqrt{g_s \left( \left(g_s n_b^2 (g_n+3 g_s)^2-q^2 (g_n+2 g_s)\right)\hbar ^4k_x^4 -2 g_s M n_b  (g_n+3 g_s) \left(4 g_s^2 n_b^2-q^2\right)\hbar ^2 k_x^2+g_s M^2 \left(q^2-4 g_s^2 n_b^2\right)^2\right)}.$ For small $k_x$, $\Gamma_k
\simeq g_sn_b  (g_n+3 g_s) \hbar ^2 k_x^2-g_s M(4 g_s^2 n^2_b-q^2)$ and  the spectrum of the two gap-less modes read 
\bea
E_{\rm m}(k_x) \simeq \pm c_{\rm m} \hbar k_x \quad  \text{and} \quad E_{\rm mp}(k_x)\simeq \pm c_{\rm mp}\hbar k_x,
\eea
where 
\bea
c_{\rm m}=\sqrt{\frac{q }{2 M}} \quad \text{and} \quad c_{\rm mp}=\sqrt{\frac{n_b  (g_n+g_s) }{M}}. 
\eea
The spectrum, and the fluctuations and magnetic currents associated with the gap-less excitations are shown in  Fig.~\ref{f:BdGspectrum}. 


%


\section{Wavefunctions of propagating MDWs}

\begin{figure*}[htp] 
	\centering
	\includegraphics[width=0.43\textwidth]{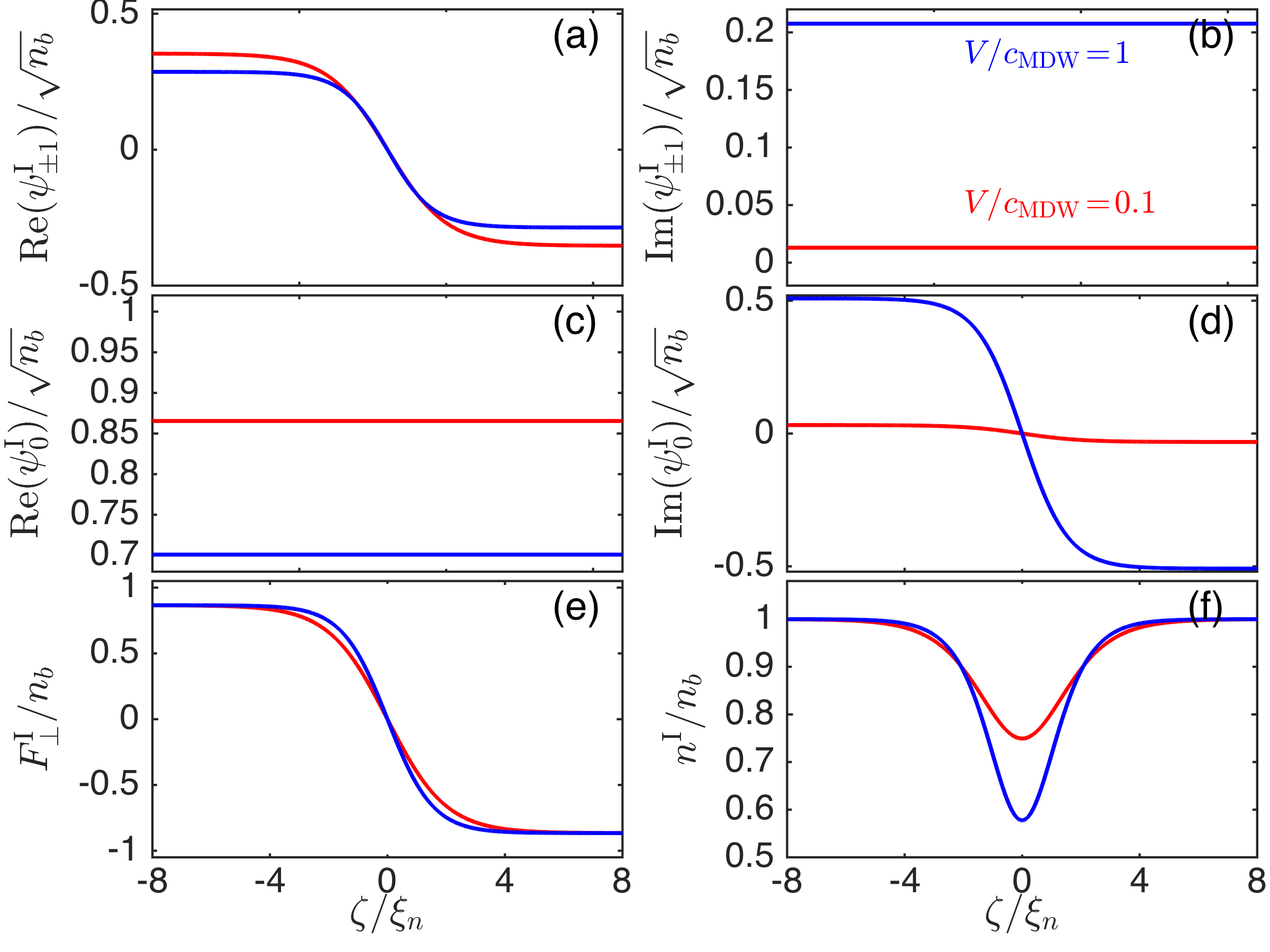}
	\hspace{1.2cm}
	\includegraphics[width=0.43\textwidth]{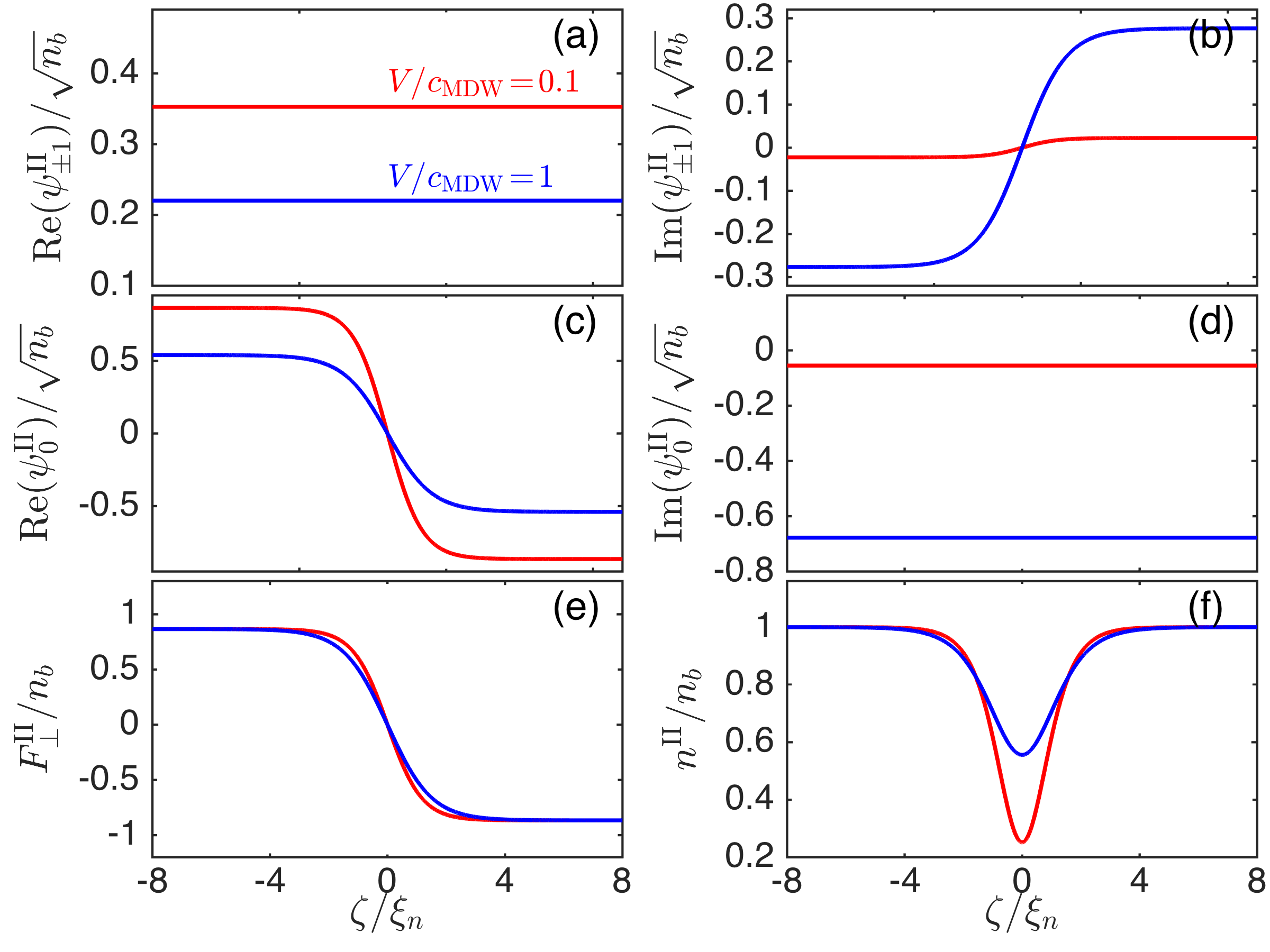}
	\caption{Wavefunctions of  type-I FDSs (left) and the type-II FDSs (right) for $V/c_{\rm FDS}=0.1$(blue) and $V/c_{\rm FDS}=1$ (red) at $g_s=-g_n/2$ and $\tilde{q}=0.5$. Here $\zeta=x-Vt$. } 
	\label{f:profile}
\end{figure*}

Fig.~\ref{f:profile} shows examples of moving FDS wavefunctions presented in Table I in the main text. The fact that $\Im (\psi^{\rm I}_{\pm 1})$ and $\Re (\psi^{\rm I}_{0})$ are constants admit exact solutions at $g_s=-g_n/2$ and $0< q<-2g_sn_b$.  Similarly, for type-II FDSs,  $\Re (\psi^{\rm II}_{\pm 1})$ and $\Im (\psi^{\rm II}_{0})$ are constants.  Away from the exact solvable regime, the constant components develop humps or dips near the domain wall core depending on the value of $g_s$.   

\section{Oscillations in a linear potential}

\subsection{Mapping to a uniform system }
It is possible to extract the FDS energy $\delta K$ from  the simulated dynamics to compare with analytical predictions which are valid for a uniform system. We construct a mapping $\psi^{m}(t) \rightarrow \tilde{\psi}^{m}(t)=\bar{\psi}^{m}_b\psi^{m}(t)/\psi^{m}_g$ for each spin state ($m=-1,0,+1$), where $\psi_g$ is the ground state in the presence of the potentials (linear+hard-wall), $\bar{\psi}_b$ is the uniform ground state with density $\bar{n}_b$. The mapped  wavefunction $\tilde{\psi}^{m}(t)$ describes a FDS in a uniform system with density $\bar{n}_b$ and the corresponding excitation energy reads 
$\delta K[\{\tilde{\psi}^{m}(t)\}]=K[\{\tilde{\psi}^{m}(t)\}]-K[\{\tilde{\psi}^{m}_b\}]$, where $K[\{\tilde{\psi}^{m}(t)\}]=\int dx \left({\cal H}[\{\tilde{\psi}^{m}(t)\}]-\mu \tilde{\psi}^{\dag}\tilde{\psi}\right)$,
$K[\{\tilde{\psi}^{m}_b\}]=\int dx \left({\cal H}[\{\tilde{\psi}^{m}_b\}]-\mu \tilde{\psi}^{\dag}_{b} \tilde{\psi}^{}_{b}\right)$, 
\bea 
{\cal H}[\psi]=\frac{\hbar^2 \left|\nabla \psi\right|^2 }{2M} +\frac{g_n}{2} |\psi^{\dag}\psi|^2+\frac{g_s}{2} |\psi^{\dag} \mathbf{S} \psi|^2 +q \psi^{\dag} S^2_z \psi 
\eea 
and $\mu=(g_n+g_s)\bar{n}_b+q/2$ is the chemical potential.

\subsection{Dynamics in components}
\begin{figure}[htp] 
	\centering
	\includegraphics[width=0.532\textwidth]{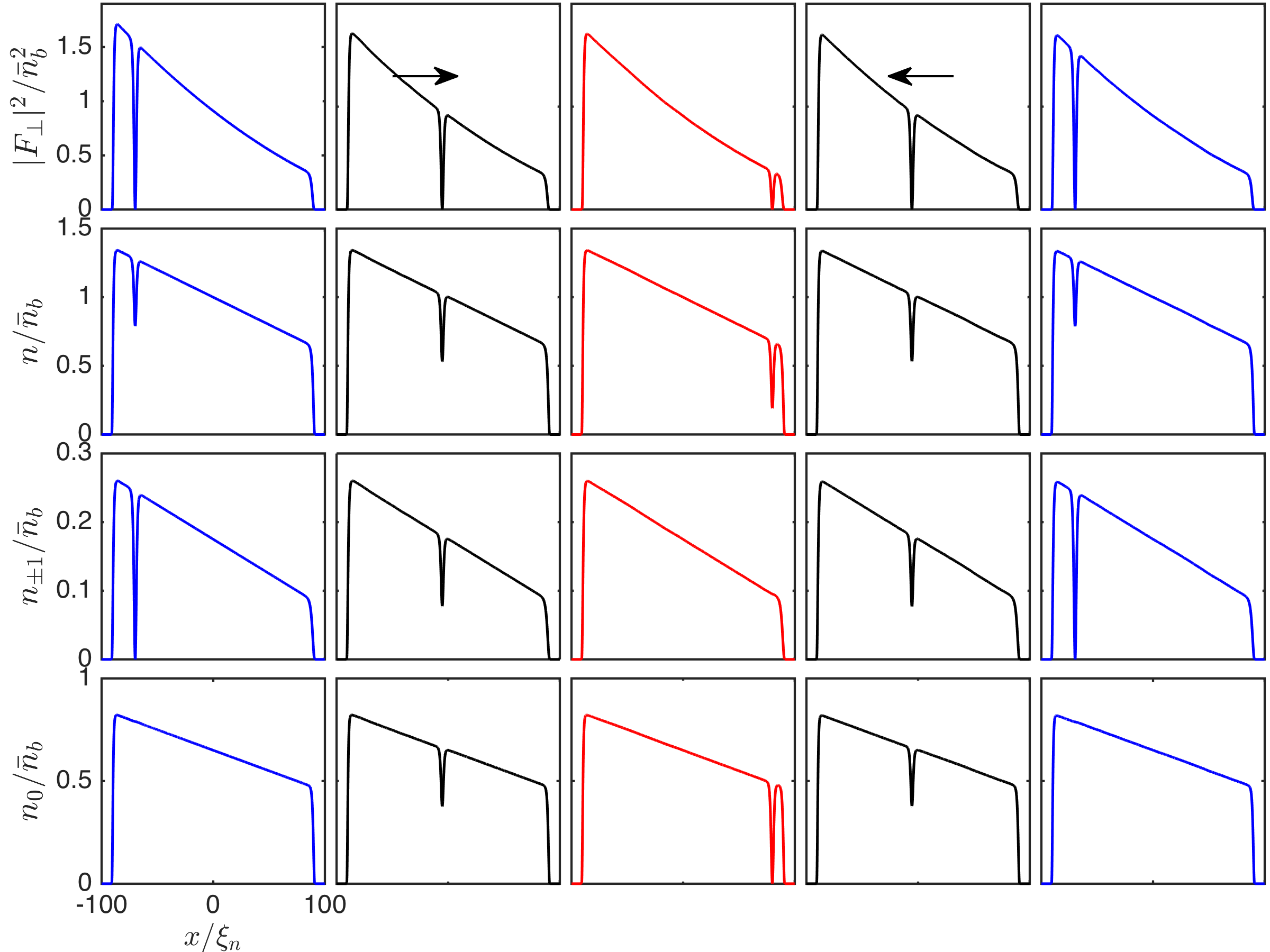}
	\caption{ A complete circle of the oscillation of a FDS in a linear potential. The parameters are the same as in Fig. 3 in the main text. The black arrows specify the evolution direction. From left to right: densities of initial state [type-I FDS with $V=0$] (blue), at the maximum velocity (black), at the turning point (red), at the negative maximum speed (back), and of the final state [returning the initial state] (blue). } 
	\label{f:oscillation4}
\end{figure}

\begin{figure}[htp] 
	\centering
	\includegraphics[width=0.446\textwidth]{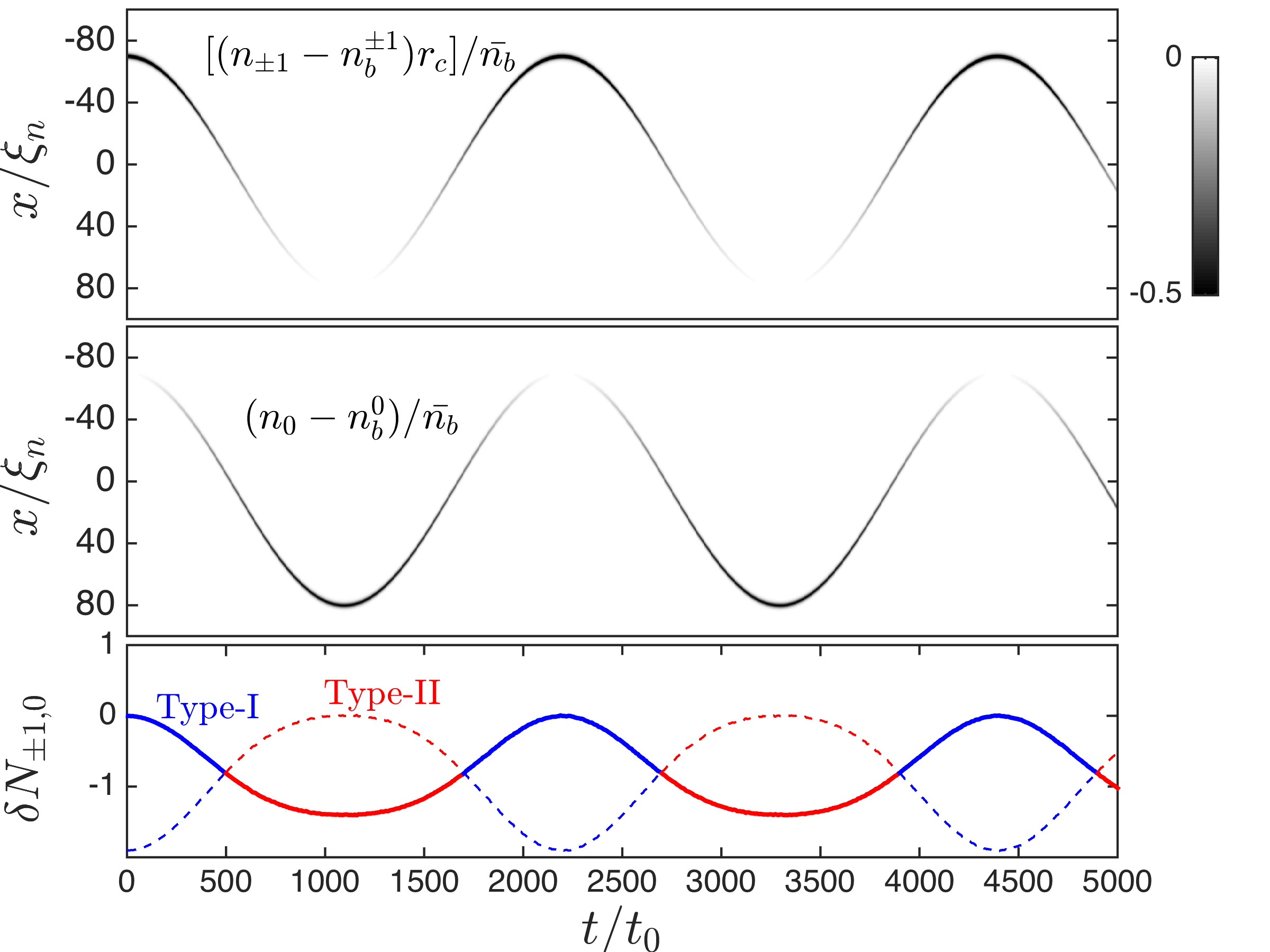}
	\caption{Internal oscillations between  $m=\pm1$ and $m=0$ spin states during the motion described in Fig.2 in the main text. Top and middle panels show component number densities (subtracting their background values) in $m=\pm 1$ and $m=0$ spin states, respectively.  Bottom panel shows oscillations of number of missing particles (due to the density dip) in $m=\pm1$ states $r_c \delta N_{\pm1}$ (solid line) and $\delta N_0$ (dashed line). At the maximum speed, $r_c \delta N_{\pm1}=\delta N_0$ and transitions between type-I (blue) and type-II (red) FDSs occur.  Here $r_c \equiv (\delta N_0/\delta N_{\pm 1})|_{V=c_{\rm FDS}}= 2\sqrt{[g_n n(x_c)+q]/[g_n n(x_c)-q]}$, $n(x_c)$ is the background density at position $x_c$ where the transition occurs.} 
	\label{f:oscillation5}
\end{figure}

Here we present further details of oscillations presented in the main text. Fig.~\ref{f:oscillation4} shows number densities and the magnetization density at different stages of the oscillation and Fig.~\ref{f:oscillation5} shows the internal dynamics of component densities.   

\subsection{Away from the exactly solvable regime}
\begin{figure*}[htp] 
	\centering
	\includegraphics[width=0.43\textwidth]{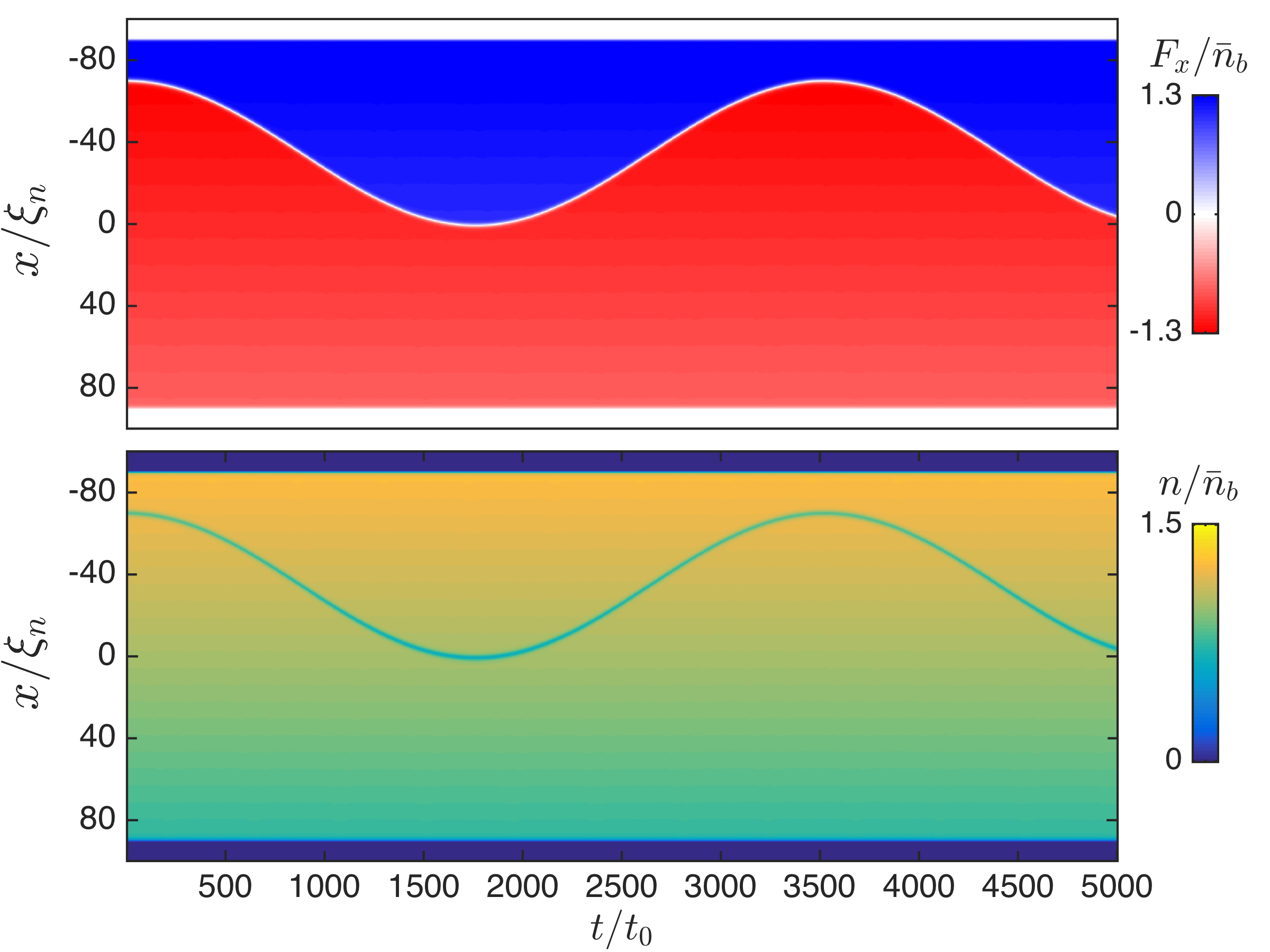}
	\hspace{1.2cm}
	\includegraphics[width=0.43\textwidth]{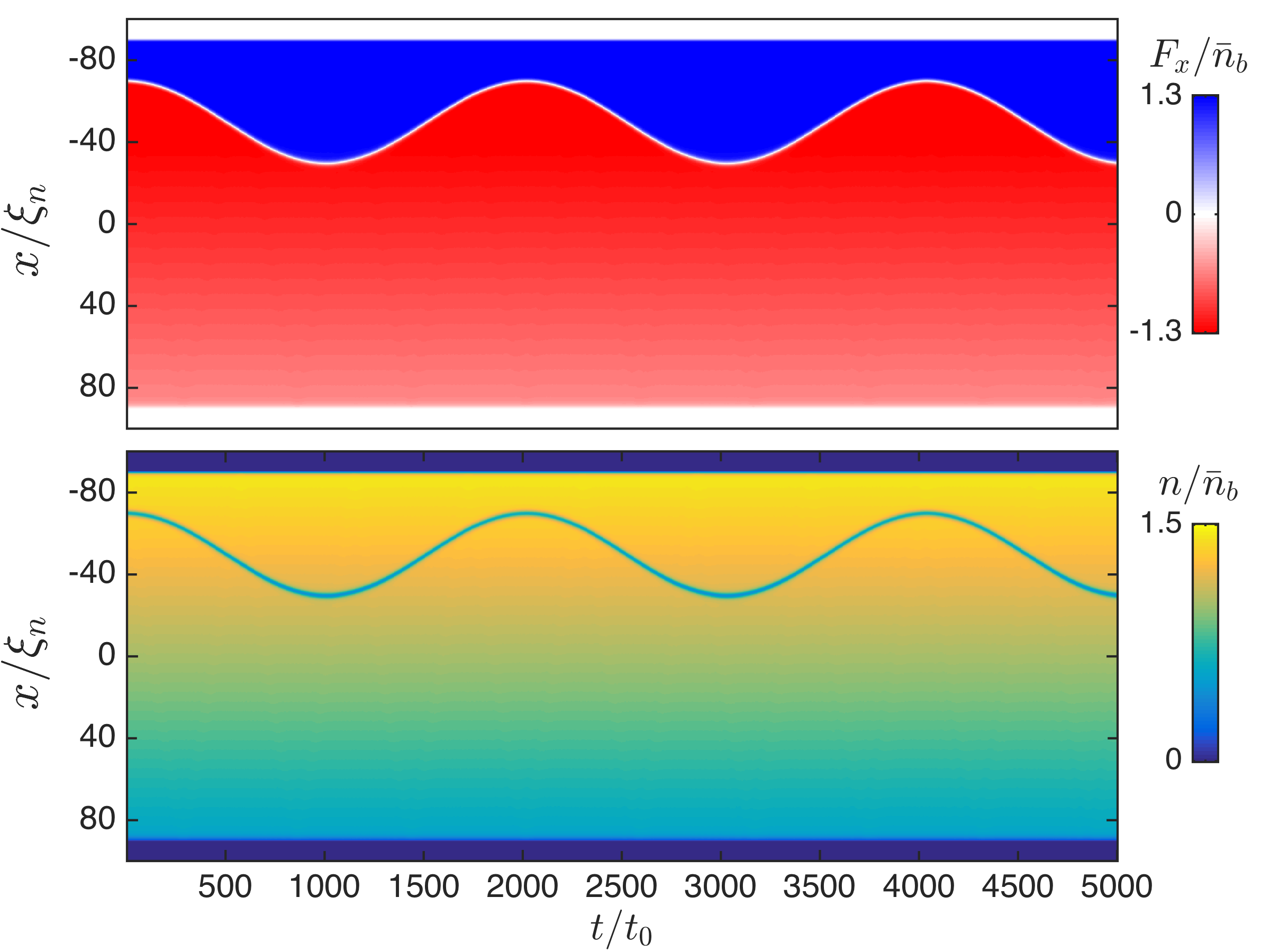}
	\caption{Oscillations away from the exactly solvable regime. The parameters are $g_s/g_n=-0.2$ and $\tilde{q}=q/(-2g_s \bar n_b)=0.1$ (left);  $g_s/g_n=-0.6$ and $\tilde{q}=q/(-2g_s \bar n_b)=0.1$ (right). The other parameters are the same as in Fig. 2 in the main text.  Here $\xi_n=\hbar/\sqrt{M g_n \bar{n}_b}$ is the density healing length, ${\bar n_b}$ is the average density, and  $t_0=\hbar/g_n {\bar n_b}$. } 
	\label{f:awayfromexact}
\end{figure*}

The properties exhibited by the exact solutions  have no qualitative change away from the exact solvable regime ($g_s=-g_n/2$, $0< q<-2 g_s n_b$). Fig.~\ref{f:awayfromexact} shows oscillatory  dynamics for $g_s/g_n=-0.2$ and $g_s/g_n=-0.6$.

\end{document}